 \documentclass[final,5p,times,twocolumn,authoryear]{elsarticle}

\usepackage{newtxtext,newtxmath}

\usepackage{xcolor}

\usepackage[T1]{fontenc}

\usepackage{graphicx}	
\usepackage{amsmath}	
\usepackage{cancel}
\usepackage[autostyle]{csquotes}
\usepackage{subfig}
\usepackage{hyperref}
\usepackage{multirow}

\journal{Astronomy $\&$ Computing}

\begin{document}

\begin{frontmatter}




\title{Accelerating radio astronomy imaging with RICK}

\author[first,second]{Emanuele De Rubeis}
\author[third,fourth]{Giovanni Lacopo}
\author[second]{Claudio Gheller}
\author[third]{Luca Tornatore}
\author[third]{Giuliano Taffoni}
\affiliation[first]{organization={Dipartimento di Fisica e Astronomia, Universit\`a di Bologna, via Gobetti 93/2, I-40129 Bologna, Italy}}
\affiliation[second]{organization={Istituto di Radioastronomia, INAF, Via Gobetti 101, 40121 Bologna, Italy}}
\affiliation[third]{organization={Astronomical Observatory of Trieste INAF, via GB Tiepolo 11, 34143 Trieste, Italy}}
\affiliation[fourth]{organization={Dipartimento di Fisica, Università degli studi di Trieste, via Alfonso Valerio 2, 34127 Trieste, Italy}}

\begin{abstract}
This paper presents an implementation of radio astronomy imaging algorithms on modern High Performance Computing (HPC) infrastructures, exploiting distributed memory parallelism and acceleration throughout multiple GPUs. Our code, called RICK (Radio Imaging Code Kernels), is capable of performing the major steps of the $w$-stacking algorithm presented in~\citet{offringa-wsclean-2014} both inter- and intra-node, and in particular has the possibility to run entirely on the GPU memory, minimising the number of data transfers between CPU and GPU. This feature, especially among multiple GPUs, is critical given the huge sizes of radio datasets involved.
\par After a detailed description of the new implementations of the code with respect to the first version presented in~\citet{gheller2023}, we analyse the performances of the code for each step involved in its execution. We also discuss the pros and cons related to an accelerated approach to this problem and its impact on the overall behaviour of the code. Such approach to the problem results in a significant improvement in terms of runtime with respect to the CPU version of the code, as long as the amount of computational resources does not exceed the one requested by the size of the problem: the code, in fact, is now limited by the communication costs, with the computation that gets heavily reduced by the capabilities of the accelerators. 
\end{abstract}


\end{frontmatter}

\section{Introduction}
\label{sec:intro}

Radio astronomy is currently witnessing a rapid increase in the volume of data that are being collected by radio-interferometers like the LOw Frequency ARray  \citep[LOFAR,][]{2013A&A...556A...2V}, MeerKAT \citep[][]{2016mks..confE...1J}, the Murchison Widefield Array \citep[MWA,][]{2010rfim.workE..16M}, the Australian Square Kilometre Array Pathfinder \citep[ASKAP,][]{2007PASA...24..174J}. These instruments can produce petabytes of data every year, precursors of what will be delivered by the Square Kilometre Array (SKA\footnote{\url{https://www.skatelescope.org/}}), expected to generate hundreds of petabytes of data each year.
\par In \citet{gheller2023} (hereafter Paper I) we have introduced a novel approach for implementing the $w$-stacking algorithm~\citep{offringa-wsclean-2014} for imaging on state-of-the-art High Performance Computing (HPC) systems, effectively exploiting heterogeneous architectures, consisting of thousands of multi-core CPUs equipped with accelerators like GPUs to enhance computational performance while minimising power consumption. 
\par Imaging is a computationally intensive step in the data processing pipeline~\citep[for an extensive introduction we refer to][and references therein]{1999ASPC..180.....T}, requiring a significant amount of memory and computing time. This is due to operations such as gridding, which involves resampling the observed data on a computational mesh, and fast Fourier transform (FFT), which converts between Fourier and real space. The computational demands increase when dealing with observations that have large fields of view, especially in current radio interferometers and at low frequencies. This is because curvature effects cannot be ignored, making the problem fully three-dimensional. In such cases, a solution is to introduce a \enquote{$w$-term} correction~\citep[see Sec.~\ref{sec:w-stacking},][]{2008ISTSP...2..647C,offringa-wsclean-2014}. The gridding, FFT, and $w$-correction steps are integrated into the so-called {\it $w$-stacking gridder} algorithm. These steps are suitable to distributed memory parallelism, exploiting parallel FFT solutions and relying on a Cartesian 3D computational mesh, that can be effectively distributed and efficiently managed across different processing units, resulting in good scalability on large HPC architectures.
\par In modern HPC systems, besides distributed processing, performance can be achieved through multi-core and accelerated (many-core) computing based on GPUs. Current trends suggest that some form of heterogeneous computing will be prevalent in emerging architectures~\citep[e.g.,][]{kekler2011}. Therefore, the ability to fully exploit new heterogeneous and many-core solutions is of paramount importance towards achieving optimal performance. In Paper I we have described the enabling of the gridding and $w$-correction algorithms to multi/many-core parallelism. We have demonstrated how utilising GPUs can significantly decrease the time to solution thanks to their outstanding performance. In addition, we have pointed out how multi/many-threads solutions allow using a smaller number of Message Passing Interface (MPI) tasks compared to a pure MPI set-up, mitigating communication-related problems. However, to fully harness power of GPU acceleration, it is critical to enable the full code for GPUs, including MPI communication. This is essential to:
\begin{itemize}
\item speed-up all algorithmic components, in particular the FFT step. The performance of the code is hindered by non-accelerated parts, which become bottlenecks;
\item avoid unnecessary data movements between the host and the device, required, in the code presented in Paper I, to implement data communication among CPUs and for the FFT transform. Efficient code performance relies heavily on minimising data movements;
\item exploit high performance interconnect available among GPUs on the same computing node.
\end{itemize}
\par In this paper, we discuss how we have tackled these challenges by exploiting compilers and libraries provided by the NVIDIA HPC Software Development Kit (SDK\footnote{\url{https://docs.nvidia.com/hpc-sdk/index.html}}). These libraries enable us to perform accelerated FFT calculations and to optimise inter-GPU MPI communication. This leads to a code that, once ones has loaded the input visibilities from the file system, can fully run on GPUs; the CPUs are used once more only at the end of the calculation, solely for the purpose of saving the final image to a file. In addition, we present a solution based on FFTW that allows the utilisation of hybrid multi-core plus MPI calculation of the Fourier transform on CPUs. This is crucial to reduce the communication between MPI tasks while also providing an efficient and portable solution that is not dependent on the NVIDIA SDK. The code, called Radio Imaging Code Kernel (RICK), is developed using the C programming language standard (with extensions to C++ only to support GPUs through CUDA). The code is publicly available\footnote{The code is publicly available here:~\url{https://github.com/ICSC-Spoke3/RICK}}, with all the details on the compilation and execution of the code for different architectures; in future releases, we plan to build a specific container for RICK, for the sake of reproducibility of our results. Alongside CUDA, in Paper I we have introduced the Open Multi-Processing (OpenMP) support for offloading gridding and w-stacking operations to accelerators, providing a portable solution even for GPU implementation. The code is compatible with various computing platforms, although optimal performance requires the availability of suitable hardware and software solutions. Parallelism has been exploited supporting multi-core processors (via OpenMP) and distributed HPC architectures (via MPI). Throughout the paper, we refer to {\it computing} or {\it processing unit} as the computing entity addressing some parts of the work. In the case of parallel work based on MPI, a computing unit is a single core (mapping to an {\it MPI task}). In the case of multi-threaded OpenMP implementation, it is a multi-core CPU. In the case of accelerated computing, it is a GPU. 
\par We present the results obtained on a state-of-the-art supercomputing platform, namely the Leonardo pre-exascale system operated by CINECA, the Italian National Supercomputing Centre, ranked as seventh in the TOP500 list\footnote{\url{https://www.top500.org}} of June 2024. CPU and GPU tests have been performed using the same code base switching between different building options by selecting the proper flags and macros in the Makefile and exploiting the NVIDIA SDK available on the system. All tests have been performed using LOFAR datasets, representative of current SKA-pathfinder radio-observations.
\par The paper is organised as follows. The methods used for performing the $w$-stacking is described in Sec.~\ref{sec:w-stacking}, together with a summary of the code presented in Paper I. In Sec.~\ref{sec:evolution} we introduce the solutions adopted for the full GPU enabling of RICK. In Sec.~\ref{section:results} the results of the performance and scalability tests are presented and discussed. Sec.~\ref{sec:reduce} is devoted to the communication and to its impact on the code. Conclusions are drawn in Sec.~\ref{sec:conclusions}.

\section{The $w$-stacking gridder}
\label{sec:w-stacking}

An interferometer measures complex visibilities $V$ related to the sky brightness distribution $I$ as:
\begin{equation}
\begin{split}
V(u,v,w) = &\int\int \frac{I(l,m)}{\sqrt{1-l^2-m^2}} \times \\ 
         &e^{-2\pi i \left(ul + vm + w(\sqrt{1-l^2-m^2}-1)\right)} dl dm,
\label{eq:visI}
\end{split}
\end{equation}
where $l,m$ are the sky coordinates while $u,v,w$ are the coordinate of the baselines in units of $\lambda$, where $w$ is chosen to be in the direction of the source. For a particular interferometer, the so-called $u,v$ plane coverage is the distribution of the baselines in $\lambda$ units as seen from the source at infinity: each projected baseline corresponds to a point of coordinates $u,v$ in the Fourier space. For small fields of view, the term $\sqrt{1-l^2-m^2}$ is close to one, and Eq.~\ref{eq:visI} is an ordinary two-dimensional Fourier transform, which, in order to speed-up the computation, is solved by using a FFT-based approach. This, however, requires mapping visibilities, that are point like data, to a regular mesh that discretizes the ($u$,$v$) space. This is accomplished by convolving the visibility data with a finite-size kernel, which converts it to a continuous function, which can then be FFT transformed.
\par When large Fields of View (FoV) are observed at once, visibility data from non-coplanar interferometric radio telescopes cannot be accurately imaged with a two-dimensional Fourier transform and the imaging algorithm needs to account for the $w$-term, which describes the deviation of the array from a plane. A possible approach to account for the $w$-term is represented by the $w$-stacking method~\citep{offringa-wsclean-2014}, in which the computational mesh has a third dimension in the $w$ direction and visibilities are mapped to the closest $w$-plane. Once gridding is completed, each $w$-plane is Fourier transformed separately, and a correction is applied as:
\begin{equation}
\begin{split}
\frac{I(l,m)\left(w_{\max} - w_{\min}\right)}{\sqrt{1-l^2-m^2}} = &\int\limits_{w_{\min}}^{w_{\max}} e^{2\pi i w(\sqrt{1-l^2-m^2}-1)} \times 
\\
&\iint V(u,v,w)  e^{2\pi i \left(ul + vm\right)} du dv dw.
\label{eq:wstacking}
\end{split}
\end{equation}
Overall, this algorithm is faster than $w$-projection~\citep{2008ISTSP...2..647C} thanks to its algorithmic characterisation, especially when the gridding is the dominating cost of the algorithm~\citep[see Tab.~1 in][]{offringa-wsclean-2014} and results in slightly lower imaging errors, making it commonly used for wide-field radio interferometers such as LOFAR or, potentially, the SKA.
\begin{figure}
\includegraphics[width=0.50\textwidth]{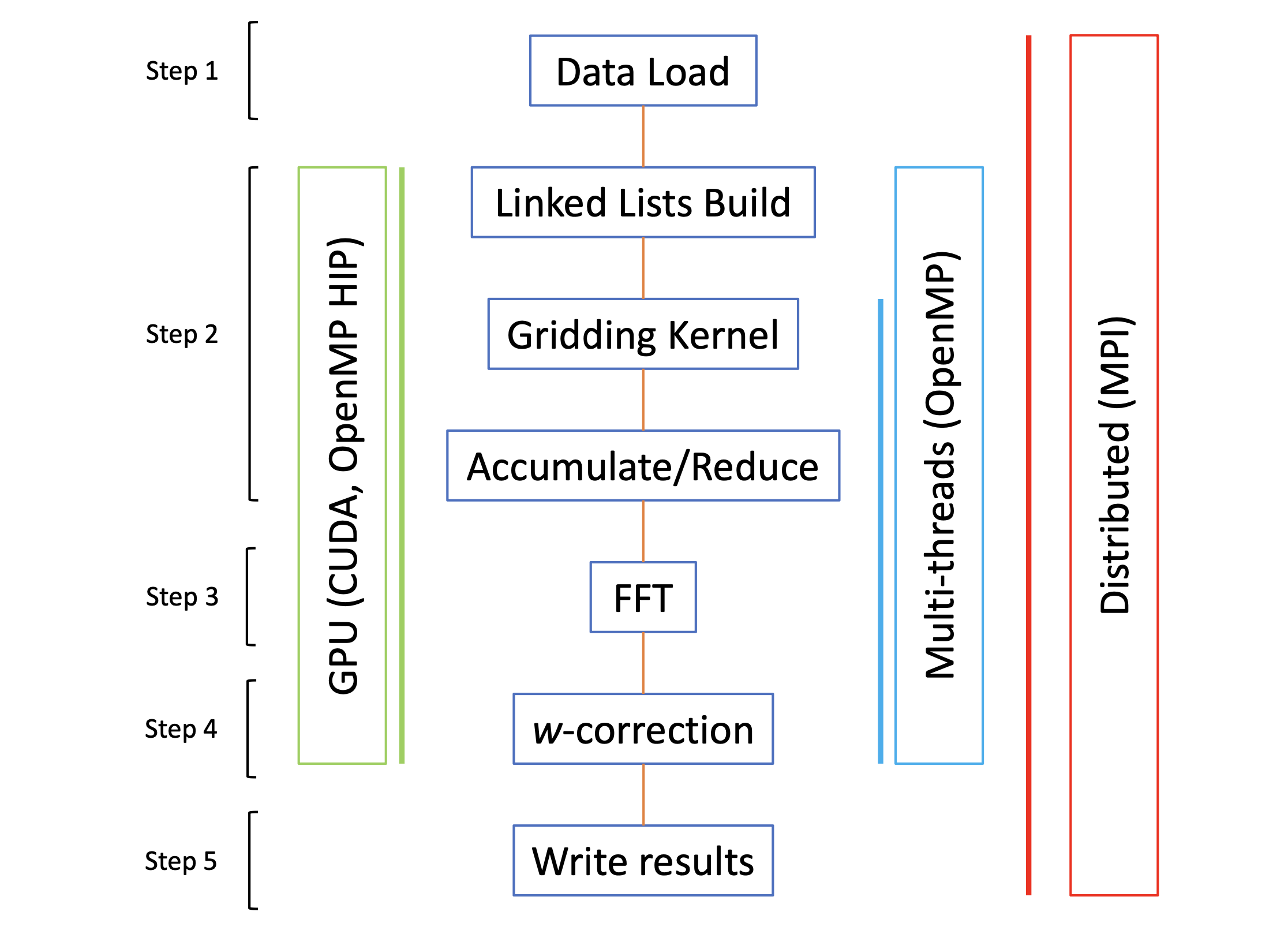}
\caption{Schematic code architecture and workflow of RICK, based on the one in~\citet{gheller2023} with the new steps that we ported on GPUs (reduce and FFT). Different kind of HPC enabling are highlighted with different colours. }
\label{fig:workflow}
\end{figure}
\par The $w$-stacking algorithm has been implemented in Paper I inside the $w$-stacking gridder, which we briefly recall hereafter. Two main data structures characterise the algorithm. The first is an unstructured dataset storing the ($u,v,w$) coordinates of the antennas array baselines at each measurement time. Each baseline has a number of associated visibilities, depending on the frequency bandwidth, the frequency resolution, the number of correlations and the total observing time. A further quantity called \textit{weight} is assigned by the correlator to each measurement: this number allows the user to adjust the balance between angular resolution, sidelobes of the beam and sensitivity~\citep{briggs1995}. Visibility data are distributed among multiple memories in time slices of equal length. The second relevant data structure is a Cartesian computational mesh of size $N_u \times N_v \times N_w$, where $N_u$, $N_v$ and $N_w$ are the number of cells in the three coordinate directions. The convolved visibilities and their FFT transformed counterpart are calculated on this mesh. The data defined on the Cartesian computational grid is partitioned among multiple tasks adopting a rectangular slab-like decomposition, assigning to each task a rectangular region of $N_{\rm{mesh}} = N_u \times N_w \times (N_v/N_{\rm{pu}})$ cells, where $N_{\rm{pu}}$ is the number of physical memories (MPI tasks for CPUs or the number of GPUs) adopted in the computation. The two data structures determine the algorithm's memory request.
\par The code consists of five main algorithmic components (shown in Fig.~\ref{fig:workflow}), each supporting different types of HPC implementations. The first component reads the data in parallel and distributes equal chunks to all the MPI tasks. If GPUs are used, the data offload from the host to the device memory completes this step. The following step performs the gridding of the visibilities. Gridding is done in successive rectangular slabs along the $v$-axis. This procedure consists of three sub-steps. In the first sub-step, an array is created for each slab, concatenating the data with $u$-$v$ coordinates inside the corresponding slab. The second sub-step is represented by the convolution with the gridding kernel:
\begin{equation}
    \tilde V(u_i,v_j,w_k) = \sum_{m \in {\rm measures}} V_m G((u_m,v_m,w_m),(u_i,v_j,w_k)),
\label{eq:convolution}
\end{equation}
where $m$ is the $m$-th measurement, $(u_m,v_m,w_m)$ are its coordinates, $(u_i,v_j,w_k)$ is a computational grid point, $V_m$ is the measured visibility and $\tilde V$ is the visibility convolved on the mesh. For the tests presented in this paper, the $G$ kernel is a Kaiser-Bessel function~\citep{jackson_91}. Once boundary data among different slabs are properly managed, this part of the computation is completely local and can be executed on multiple cores or accelerators. This step is finalised by the exchange of information among different computing units, necessary to accomplish the calculation of gridded visibilities on each slab. This is a critical operation, whose detailed description is given in Paper I.
\par The third algorithmic component performs the FFT of the gridded data, producing the real space image. Next, we apply the phase shift and reduce the $w$-planes to obtain the final image (Step 4). The fifth and final step writes the final images exploiting parallel I/O.

\section{Multi/Many Core Based HPC Architectures Support}
\label{sec:evolution}

In this Section, we discuss the newly implemented features of RICK, including the support for multi-many threads FFT, the integration of MPI based direct GPU-GPU communication and the introduction of an optimised hybrid shared-distributed memory to perform the reduce operation in substitution of the standard MPI library calls.

\subsection{Parallel FFT on the GPU}
\label{sec:fft_cuda}
Running the FFT directly on the GPUs represents a pivotal step for the RICK code since it reduces the computational cost of this operation and removes the data movement associated with it (see Paper I).
\par We have adopted the NVIDIA CUDA FFT library (cuFFT\footnote{\url{https://docs.nvidia.com/cuda/cufft/contents.html}}), which provides an interface for computing FFTs on an NVIDIA GPU. In particular, given our distributed memory parallelism approach, we exploited the distributed version of the cuFFT, namely the cuFFT Multi-process library (cuFFTMp\footnote{\url{https://docs.nvidia.com/hpc-sdk/cufftmp/index.html}}). This is a multi-node, multi-process extension to cuFFT that supports multiple GPUs across multiple nodes, a key feature given that large size problems hardly fit into a single GPU memory. At the time of writing, cuFFTMp is the only library capable of solving the FFT problem among distributed GPUs. Specifically, we have used version 11.0.14 of the cuFFTMp, included in version 23.11 of the NVIDIA HPC SDK toolkit. The cuFFTMp library uses NVSHMEM\footnote{\url{https://developer.nvidia.com/nvshmem}}, a communication library based on the OpenSHMEM standard that creates a global address space that includes the memory of all GPUs in the cluster. The main steps of the cuFFTMp implementation of the 2D FFT on NVIDIA GPUs are listed hereafter.
\begin{itemize}
    \item A plan for 2D complex-to-complex FFT is created, which contains all information necessary to compute the transform, including the pointers to the input and output arrays.
    \item For each $w$ plan, distributed data are copied to the respective \textit{descriptor}, that represents an ad-hoc structure for data that have to be transformed. 
    \item The previously-initialised plan is executed doing a 2D inverse Fourier transform.
    \item Data are redistributed to the original order with a DEVICE\_TO\_DEVICE copy, because after the FFT they are distributed in a permuted order.
    \item Finally, transformed data are brought back from descriptors and written into a new device pointer, always distributed among multiple GPUs, which is then used for the following steps of the code.
\end{itemize}
This operation is executed for each $w$ plan, as required by the $w$-stacking algorithm. Within the loop over $w$, we need to allocate and free device memory for the first descriptor used by the cuFFTMp, that is the one used for the execution of the FFT plan. A second descriptor is instead used only to redistribute data in the original order, so it can be allocated and freed just once outside the $w$-layers loop. The number of layers to be transformed has a significant impact on the overall performance of this step. Also the initialisation of the library, in particular with the creation of the plan at the beginning of the FFT step, impacts the computing time. This will be further discussed in Sec.~\ref{sec:strong_scaling}.
\par Having data resident on the GPU memory, we managed to use a CUDA kernel to write the to-be-transformed input array into a \textit{cufftDoubleComplex} data type, dividing the real and imaginary parts, and copying it directly within the descriptor, speeding up its creation and optimising the requirement of adopting this kind of data structure. We also used a CUDA kernel to write FFT transformed data back to the right pointer on the GPU which is then used for the following the $w$-correction step. This fully eliminates the communication between host and device, hence preventing the need for data transfer between the CPUs for every process and $w$-plan.

\subsection{Hybrid CPU-based FFT}
\label{sec:fft_hybrid}
In the code presented in Paper I, the FFT algorithm could exploit only the MPI implementation (FFTW-MPI\footnote{\url{https://fftw.org/doc/Distributed\_002dmemory-FFTW-with-MPI.html}}). When GPUs are used, each of them is assigned to a single MPI task, running on a single CPU core. These cores are the only ones that can contribute to a pure MPI FFTW calculation: therefore, a significant fraction of the node's computational capacity is wasted. In order to overcome this limitation, we have exploited the MPI+OpenMP parallelism, using the Hybrid FFTW\footnote{\url{https://www.fftw.org/fftw3\_doc/Combining-MPI-and-Threads.html}} which allows the combination of MPI tasks and OpenMP threads to fully utilise the CPU as follows.
\begin{itemize}
    \item First, data are distributed among all the MPI tasks as in the FFTW-MPI implementation.
    \item Then, each MPI task spawns a certain number of OpenMP threads for a further workload distribution.
\end{itemize}
This allows exploiting all the available hardware: in the previous example, the hybrid FFTW runs on 4 MPI tasks, each task spawning 8 threads, effectively maximising the CPU performance. 
In addition, when all the cores of each CPU are used, the hybrid implementation is expected to scale better compared to the pure MPI one, since the number of MPI tasks performing communication is reduced, leading to a lower communication overload. Communication is further optimised adopting the strategies discussed in Sec.~\ref{sec:communication}.

\subsection{Communication}
\label{sec:communication}

The reduce operation consists in summing contributions gathered from all the MPI tasks in the memory of a target processor or GPU. This can introduce a significant overhead even on a single node when the number of MPI tasks per node is large or the problem size increases (see Paper I for details).
\par To reduce such overhead, we have adopted two different solutions. First, we support direct GPU-GPU communication, which allows $i)$ exploiting the NVlink NVIDIA high-speed interconnect within a node, $ii)$ avoiding the CPU-GPU data transfer necessary for standard MPI communication. Second, we exploit MPI and OpenMP to combine shared and distributed memory data access in order to $i)$ optimise the access to local data, $ii)$ minimise the number of MPI tasks and related message passing communication overheads. 

\smallskip
\noindent{\bf GPU-based Reduce.}
Direct GPU-GPU communication has been implemented by exploiting the NVLink high-speed interconnect for NVIDIA and InfinityFabric, the corresponding technology available for AMD architectures. NVLink is a wire-based communications protocol that can be used for data and control code transfers in processor systems between CPUs and GPUs and solely between GPUs. The Leonardo Booster is equipped with NVLink 3.0, which provides a transfer rate of 50 Gbit/s and a bandwidth per GPU of 600 GB/s. 
\par We refer to NVIDIA Collective Communication Library (NCCL\footnote{\url{https://developer.nvidia.com/nccl}}) which implements the reduce operation as a ring intra-node, and an inter-node ring, when GPUs assigned to the main tasks communicate with Remote Direct Memory Access (RDMA) with GPUs in different nodes without passing through the CPUs~\citep{Li2020}. A similar solution can be adopted for AMD GPUs thanks to ROCm Collective Communication Library (RCCL\footnote{\url{https://rocm.docs.amd.com/projects/rccl/en/latest/}}), by simply replacing CUDA calls with HIP calls.

\smallskip
\noindent{\bf Hybrid CPU-based Reduce.}
To mitigate the impact of reduce operations, we have designed a hybrid reduce technique which combines MPI and OpenMP parallelism and exploits the Non-Uniform Memory Access (NUMA) topology. This hybrid implementation works as follows: as first, two MPI communicators are built: $1)$ an intra-node communicator, where only the MPI ranks pertaining to each computing node are included and are ranked from $0$ to $P-1$, with $P$ being the number of MPI ranks on every node, $2)$ an inter-node communicator that groups all ranks $0$ in the intra-node communicators. In communicator $1$, each MPI task knows its siblings and exchanges data using shared-memory windows with a ring algorithm. Each MPI task spawns two threads. Thread 1 is in charge of calling the shared-memory reduce function with a ring fashion involving all the other tasks in the intra-node communicator with its siblings, while thread 0 manages the MPI communications among the nodes. It is important to say that although each task needs at least two threads, only task 0 (i.e. the master) is actually involved in the MPI communications. So in the current implementation all the other tasks spawn \enquote{silent} threads. In a future implementation we plan to fully occupy the CPU cores with OpenMP and let all the threads participate to the summation. When the shared-memory reduce is done, task 0 gathers the result: this is the total summation in case of a single node run, and it is a partial summation in multi-node cases. The partial sums on each node are finally added using an MPI Ireduce call among the tasks in communicator $2$, resulting in the overall sum. Hereafter, when we discuss results about hybrid MPI+OpenMP tests, we assume that the hybrid reduce is used; a future paper will be dedicated on the development of this reduce.

\section{Performance Analysis}
\label{section:results}

To evaluate the performance of RICK, we run a number of tests exploiting the Leonardo HPC architecture, available at the CINECA supercomputing centre. CINECA is the Italian national HPC facility and the Leonardo Booster system is made of 3456, 32-cores, Intel Xeon Platinum 8358 CPU nodes equipped with 4 NVIDIA Tesla Ampere 100 GPUs. Each CPU node has 512 GB of DDR4 memory, the memory of the GPU is a 64 GB HBM2. The CPU and the GPU are interfaced by a PCIe Gen4 interconnect, while the 4 GPUs per node adopt NVLink 3.0. The system interconnect is a Nvidia Mellanox network, with Dragon Fly+, capable of a maximum bandwidth of 200Gbit/s between each pair of nodes. The systems deploys a Lustre parallel file system as working storage area. The source code has been compiled using the GCC and the NVCC compilers and OpenMPI library v4.1.4. The only dependency of the code is on the FFTW3 library. The Makefile requires only few adjustments of several variables (e.g. the path to FFTW or to the CUDA libraries, the gridding kernel, the enabling of GPUs) to compile the code. The building procedure requires only a few seconds. The GPU tests have been performed using CUDA version 12.1 and NVIDIA HPC SDK version 23.11: this suite includes NCCL version 2.18.5, NVSHMEM version 2.10.1, and cuFFTMp version 11.0.14. 
\par Three different input data configurations have been used for the tests, namely the {\it Small}, \textit{Intermediate} and {\it Large}, addressing different test sizes. The first can run on a single CPU/GPU up to a small number of computing units. This allows for testing the performance of each unit and comparing the scaling within a node and between nodes. The second encompasses a number of intermediate computational configurations, that can be frequently expected for radio astronomical data. The third focuses on extreme cases, for which large images have to be generated and hundreds of computing units are required to ensure that enough memory is available to manage the computational mesh. The input data comes from two different radio measurement sets (MS) from LOFAR 8 hours, dual inner mode, observations in high band antenna (HBA). One of the MSs is only constituted by the Dutch stations (LOFAR Dutch), the other using the full International LOFAR Telescope (LOFAR ILT). The main characteristics of the two observations, together with the size of the computational mesh adopted in the three test cases, are presented in Tab.~\ref{table:observation} and~\ref{table:datasets}. The dataset for the \textit{Small} tests is a single sub-band of 2 MHz of the LOFAR Dutch observation at 150 MHz. The \textit{Intermediate} and \textit{Large} tests used instead the same LOFAR ILT MS, with the only difference residing in the size of the computational mesh. 

\begin{table}
\begin{center}
\centering \tabcolsep 1pt
\begin{tabular}{l|l|l}
\hline
 & LOFAR Dutch & LOFAR ILT\\ 
\hline
RA  & 15:58:18.96 & 17:12:50.04\\
Dec & +27.29.19.20 & +64.03.10.60\\
Observation time & 8 hrs & 8 hrs\\
Integration time & 4sec & 4 sec\\
N. antennas & 62 & 72 \\
Bandwidth & 1953.1 kHz & 1953.1 kHz\\
Reference frequency & 150.5 MHz & 144.6 MHz\\
N. sub-band & 1 & 1\\
N. of channels per sub-band & 20 & 160\\
Channel width & 97.656 kHz & 12.207 kHz\\
Project code & LC14\_018 & LT16\_005\\
Principal Investigator & F. Vazza & A. Botteon\\
\hline
\end{tabular}
\end{center}
\caption{Main characteristics of the LOFAR HBA datasets used for the \textit{Small} (LOFAR Dutch) and \textit{Intermediate}/\textit{Large} (LOFAR-ILT) configuration test.}
\label{table:observation}
\end{table}

\begin{table}
\begin{center}
\centering \tabcolsep 2pt
\begin{tabular}{l|l|l|l}
\hline
    & Small & Intermediate & Large \\ 
\hline
N. visibilities (approx)  & 0.54$\times$10$^9$ & 47.05$\times$10$^9$ & 47.05$\times$10$^9$ \\
Input data size (GB) & 4.4 & 533 & 533 \\
$N_u = N_v \times N_w$ & $4096^2\times$16 & $16384^2 \times 32 $ & $65536^2\times$32 \\
Mesh size (GB)  & 10.81 & 258.81 &4098.81 \\
\hline
\end{tabular}
\end{center}
\caption{Configuration and computational mesh used in the \textit{Small}, \textit{Intermediate} and \textit{Large} tests. The mesh size takes into account the total amount of memory required for the real and imaginary part depending on the size of the grid.}
\label{table:datasets}
\end{table}
\par For each test, we report the following wall-clock time measurements related to specific code components:
\begin{itemize}
    \item the gridding time, which is the time required for the gridding operation on the visibilities;
    \item the reduce time, related to the reduce operation;
    \item the FFT time, which is the time required for inverse Fourier transform of the gridded visibilities;
    \item the $w$-correction time, the last step of the code before writing the final outputs;
    \item the total parallel time, which is the total execution time of the code neglecting the initial setup time, which involves mainly I/O data load operations and the writing of the final image.
\end{itemize}

\begin{figure*}[t]
\centering
 \subfloat{\includegraphics[width=\columnwidth]{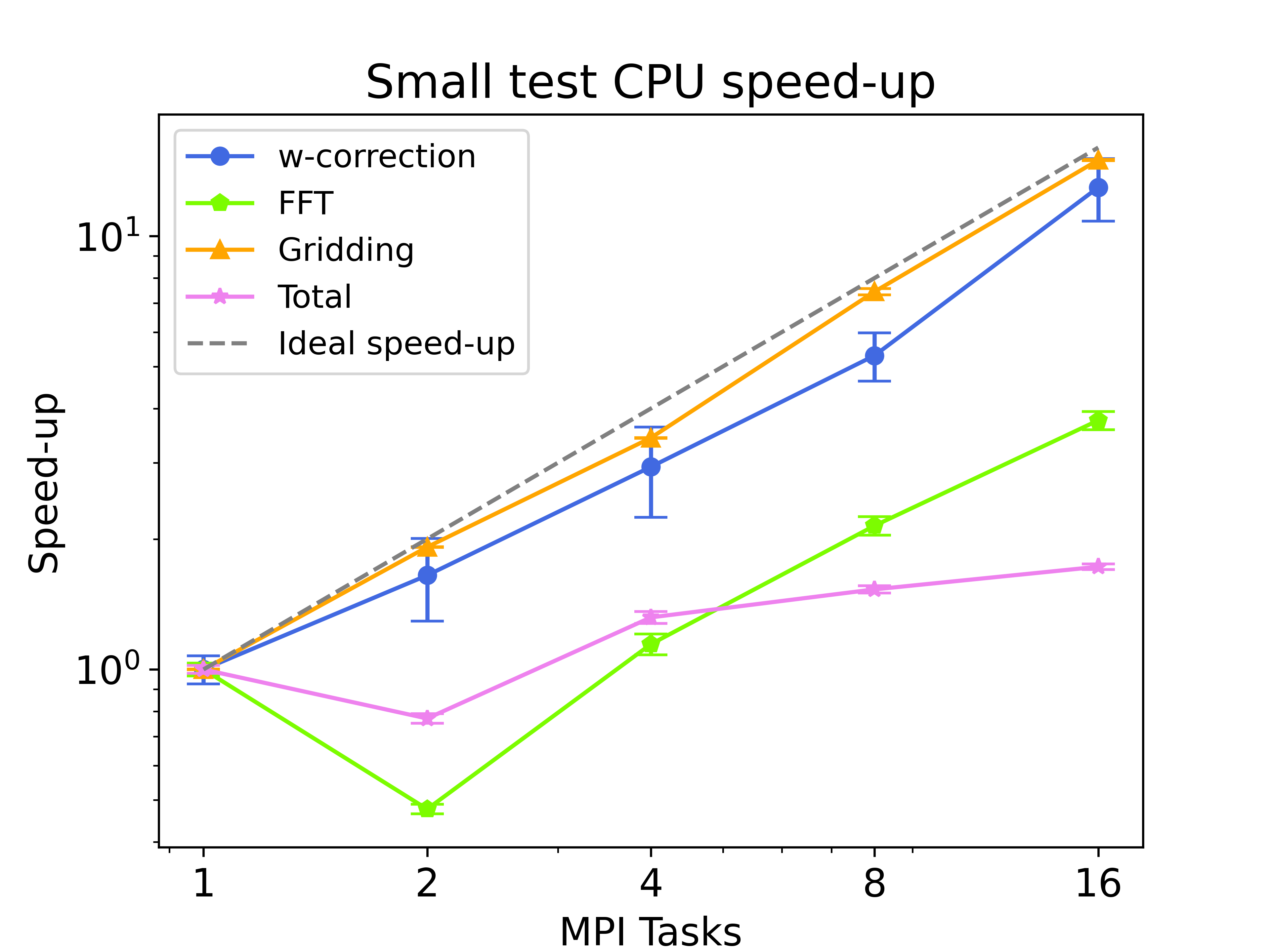}}
 \subfloat{\includegraphics[width=\columnwidth]{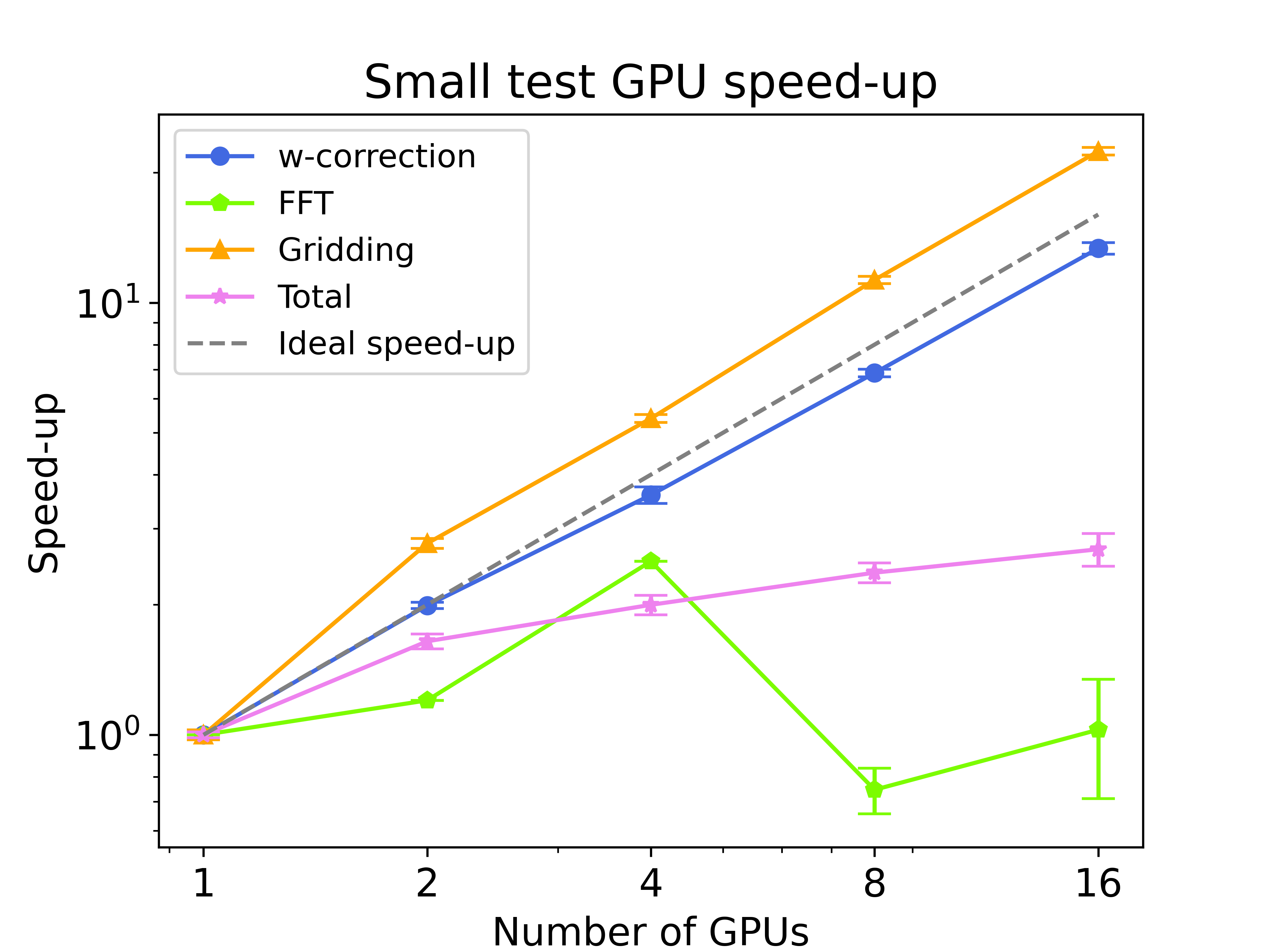}}\\
 \subfloat{\includegraphics[width=\columnwidth]{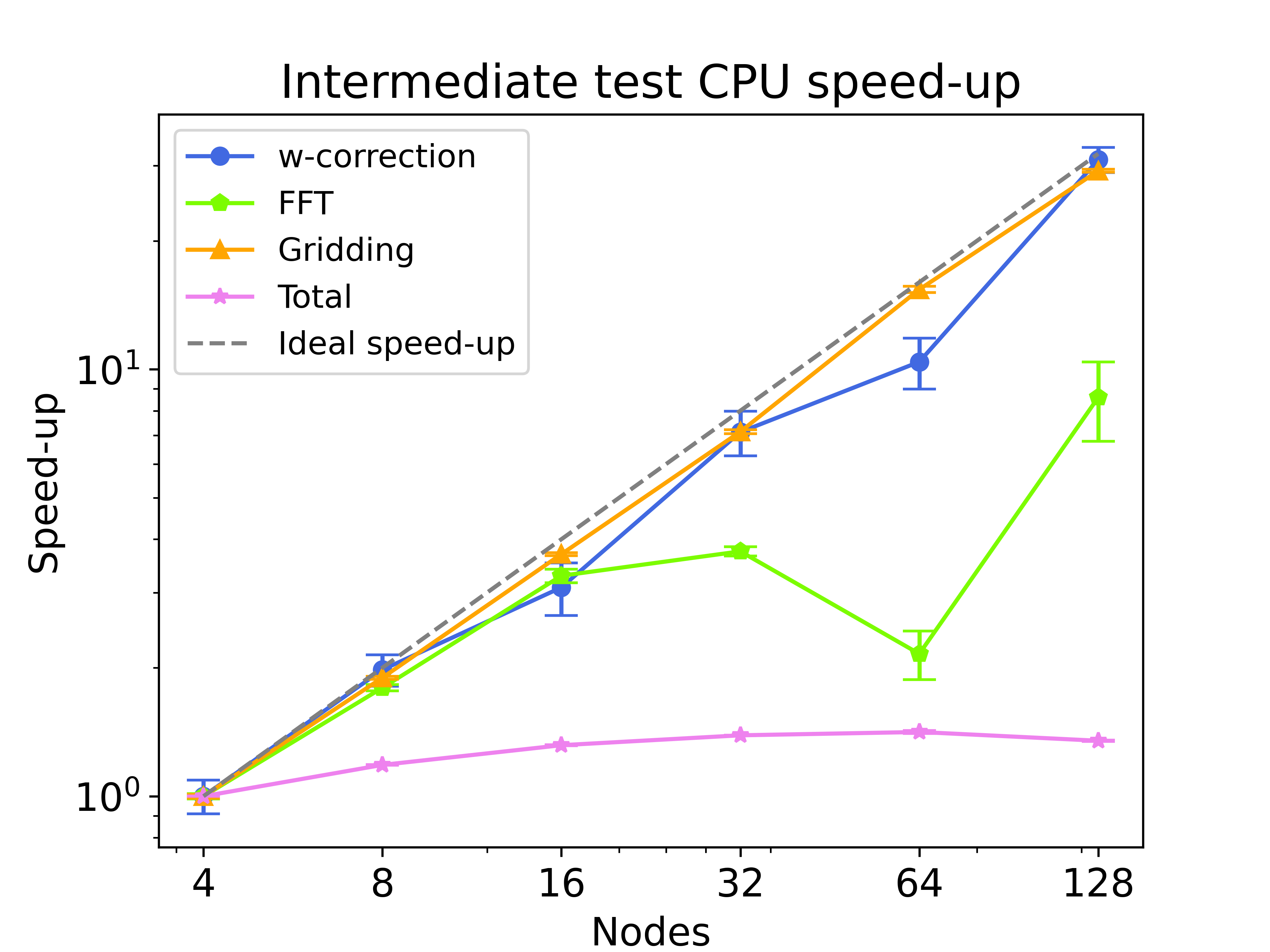}}
 \subfloat{\includegraphics[width=\columnwidth]{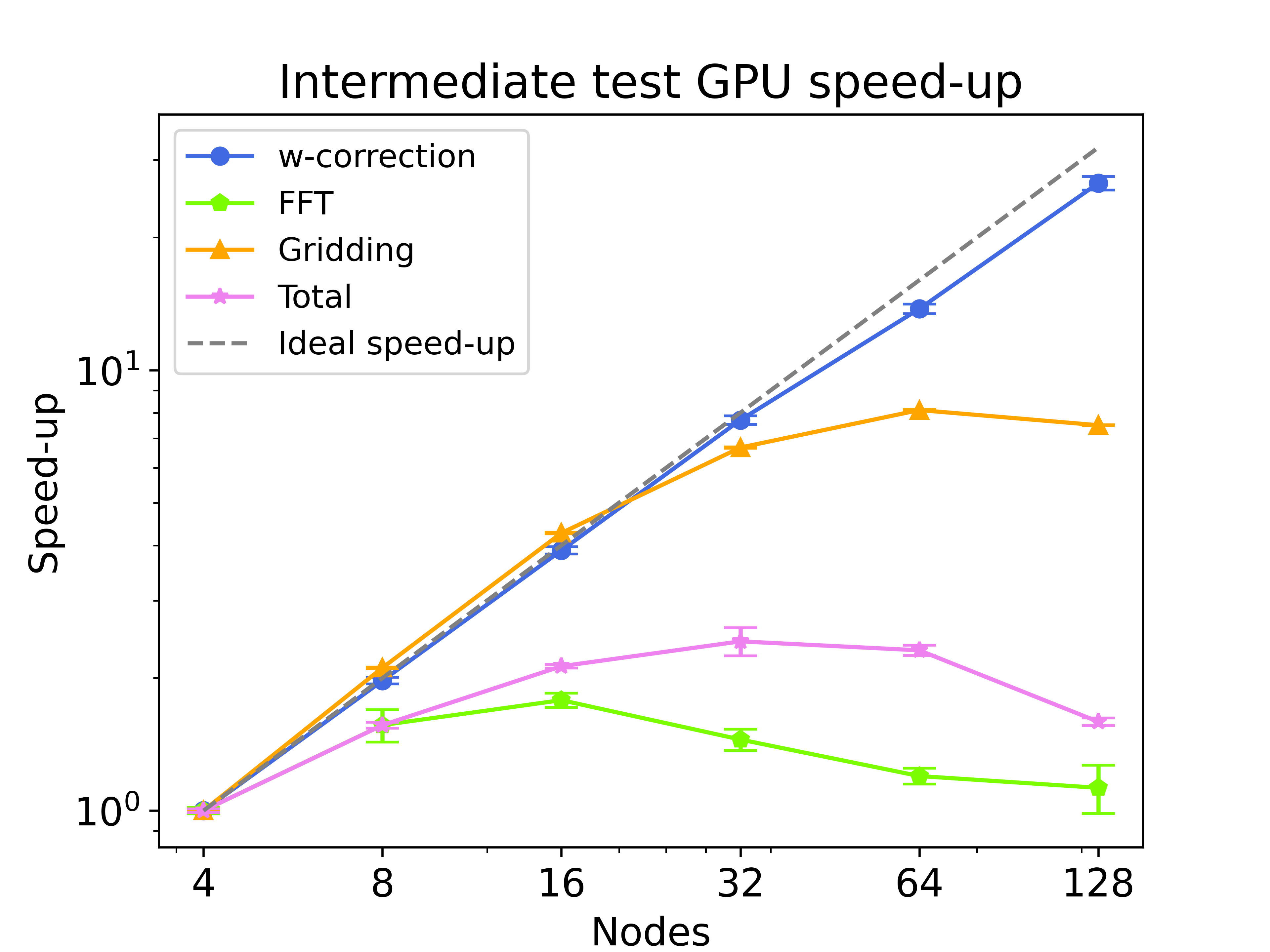}}
 \caption{Speed-up results for each step of the code for both full CPU and full GPU code, using \textit{Small} (top) and \textit{Intermediate} (bottom) dataset. The dashed grey line represents the ideal scaling trend. For the \textit{Intermediate} test, the number of resources per node is of 4 MPI tasks and 8 OpenMP threads for the full CPU and 4 GPUs and 8 OpenMP threads for the full GPU test.}
 \label{fig:speedup_plots}
\end{figure*}
In order to assess the performance of the different code components, we have measured the wall-clock times they took to complete the different tests and we have discussed the best performance for each configuration.
We have then analysed the code scalability, that can be defined as the ability to handle more work as the size of the computing system or of the application grows. Scalability has been assessed in terms of strong and weak scaling.  The former measures the performance of the code keeping the data and the mesh size constant and progressively increasing the adopted computing units. Ideally, the code execution time should scale linearly with the number of computing units (i.e. the time should halve if the number of computing units doubles). However, various factors can impact such an ideal behaviour, leading to a performance degradation with increasing the number of units. Therefore, strong scalability indicates the gain the user can expect increasing the number of computing units working cooperatively.
\par The strong scalability can be measured by the \textit{speed-up} parameter, calculated as
\begin{equation}
\label{eq:speedup}
    S = \frac{T_{1}}{T_{N_{\rm{p}}}}
\end{equation}
where $T_{N_{\rm{p}}}$ is the wall clock time measured using $N_{\rm{p}}$ computing units and $T_{1}$ is the corresponding time measured using a baseline configuration (e.g. a single computing unit).
\par In the case of weak scalability, both the number of processing units and the problem size are progressively increased, resulting in a constant memory and work load per processing unit.

\subsection{Code performance}
\label{sec:best_performance}

\begin{table*}[t]
\begin{center}
\centering \tabcolsep 1pt
\begin{tabular}{l|c|c|c|c|c|c|c|c}
\hline
& Nodes & MPI(threads) & GPUs & Gridding (s) & Reduce (s) & FFT (s) & $w$-correction (s) & Total (s)\\ 
\hline
\textit{Small} CPU & 4 & 16 (8) & 0 & 0.304$\pm$0.001 & 4.351$\pm$0.002 & 0.676$\pm$0.028 & 0.179$\pm$0.028 & 5.636$\pm$0.039\\
\hline
\textit{Small} GPU & 4 & 16 (8) & 16 & 0.104$\pm$0.001 & 0.207$\pm$0.001 & 0.615$\pm$0.04 & 0.010$\pm$0.001 & 1.187$\pm$0.048\\
\hline
\hline
\textit{Intermediate} CPU & 64 & 256 (8) & 0 & 1.719$\pm$0.026 & 138.352$\pm$0.064 & 9.646$\pm$1.253 & 0.419$\pm$0.050 & 150.944$\pm$1.256\\
\hline
\textit{Intermediate} GPU & 32 & 128 (8) & 0 & 1.244$\pm$0.026 & 10.132$\pm$0.436 & 1.803$\pm$0.058 & 0.021$\pm$0.001 & 15.965$\pm$0.440\\
\end{tabular}
\end{center}
\caption{Best performance times for \textit{Small} and \textit{Intermediate} tests both full CPU and full GPU.}
\label{table:best_times}
\end{table*}

In this section, we provide the results for each configuration to achieve the following objectives: {\it i)} inform the user about the expected runtime of the code in various setups; {\it ii)} compare the CPU and GPU time to solution. For each configuration, we showcase the best possible result achieved by varying the number of processing units. The timings encompass all the various stages involved in processing the data, including the time spent for communication and for a minor algorithmic component that has not yet been parallelised. Data reading and writing is instead not addressed here. Additional information on the single CPU/GPU performance can be found in Paper I. 
\par Best performance is presented in Tab.~\ref{table:best_times} for the \textit{Small} and \textit{Intermediate} tests, while for the \textit{Large} test it corresponds to the set-up with 64 and 32 nodes for the CPU and GPU cases respectively, reported in Tab.~\ref{table:large_times}. 
\par For the \textit{Small} configuration the best result is obtained using 4 nodes for both the CPU and the GPU tests. Both tests run with 4 MPI tasks per node, the former spawning 8 OpenMP threads per task, and latter assigning one GPU per MPI task. The total time to solution is around 5.6 seconds for the CPU setup. The GPU setup runs around 4.5 times faster, in about 1.2 seconds. The breakdown in the different algorithmic components shows that gridding and $w$-correction are around 3 and 18 times faster on the GPU respectively. The FFT times are instead comparable. The big difference emerge for the MPI reduce time which is more than 20 times faster for the GPU setup. We will discuss in details the impact of the reduce in the following sections.
\par The \textit{Intermediate} configuration show the best performance on 64 and on 32 nodes for the CPU and GPU setups respectively. Overall, the GPU setup takes around 16 seconds to complete, almost 10 times faster than the CPU one. In both setups the majority of the time is spent in the reduce operation, while the FFT becomes much faster on the GPU, thanks to the the parallel cuFFTMp implementation, much more efficient than the hybrid FFTW.
\par For the \textit{Large} tests, the GPU setup runs on 32 nodes around 150 times faster than the CPU one, achieved on 64 nodes. Such result is mainly due to the huge difference in the reduce time. Once more, a dedicated discussion on this aspect is provided below. The breakdown shows that gridding times on GPU and CPU setups are similar, with CPU time faster around 1.4 times than GPU one. We will discuss the reason of this behaviour in the next section. The breakdown shows that FFT times on GPU setup are much faster than in \textit{Intermediate} tests, around 25 times compared to CPU setup.

\subsection{Strong scaling tests}
\label{sec:strong_scaling}
Strong scaling results are shown in Fig.~\ref{fig:speedup_plots}, where the speed-up is measured as a function of the number of computing units for the \textit{Small} and \textit{Intermediate} configurations. In Tab.~\ref{table:large_times} we report the execution times for the \textit{Large} test. The tests have been conducted on a progressively larger number of computing units, varying in a range that depends on the test size. The minimum number of computing units is set so that the problem can be fit within the available memory. The maximum number of computing units is chosen to guarantee that the computing time is higher than the communication time, with the exception of the \textit{Large} test, which will be discussed separately
\par The \textit{Small} tests have been performed using up to 4 nodes with 4 MPI tasks/node for the MPI runs (up to 16 MPI tasks in total) or 4 GPUs/node for the GPU runs (with 1 MPI task associated to each GPU), with 8 OpenMP threads. \textit{Intermediate} tests range between 4 and 128 nodes, with a fixed number of 8 OpenMP threads and 4 MPI tasks (or GPUs) per node. The \textit{Large} tests required at least 32 nodes because of the large memory request. Computing resources increased progressively up to 128 nodes, which means a total of 4096 MPI tasks for pure CPU runs or 512 GPUs for the CUDA runs.
\par \noindent{\bf Gridding time.} For the GPU version of the gridding algorithm, the  timings account also for the offload time of visibilities and relative weights from CPU to GPU through asynchronous memory copies. The gridding kernel is called iteratively, once per each mesh sector (see Paper I for details), so we report here the sum of the times spent for each individual sector (namely the total gridding time).
\par For the \textit{Small} tests, Fig.~\ref{fig:speedup_plots} top panels, we first perform a baseline test using 1 MPI task and 8 OpenMP threads. All the other CPU runs double each time the number of MPI tasks, keeping 8 OpenMP threads per task. We notice the almost perfectly linear scaling of the gridding over a single node, ranging from 1 to 4 MPI tasks. Since each node has 4 GPUs, we have not explored other hybrid single node configurations any further. The tests with $8$ and $16$ MPI tasks have been performed using $2$ and $4$ computing nodes, respectively, with 4 MPI tasks per node and 8 threads. The GPU tests for the \textit{Small} dataset show a good scaling starting from $1$ GPU up to $16$ GPUs distributed again into $4$ Leonardo booster nodes. GPUs result to be $\simeq 2-3$ times faster than CPUs (see Tab.~\ref{table:best_times}).
\par For the \textit{Intermediate} strong scaling tests (Fig.~\ref{fig:speedup_plots}, lower panels), we measure the speed-up of the hybrid MPI+OpenMP implementation for CPUs (bottom left panel), observing that gridding scales almost ideally even up to 128 nodes. In this case a thread synchronisation is needed because different threads of each MPI task have to perform gridding in concurrent regions of the mesh. Anyway, in Intel CPUs, this multi-threading synchronisation, which can lead to an overhead, does not impact on the code performance and the algorithm scales linearly. For the GPU case (bottom right panel), we observe that the gridding shows an ideal scaling only up to 16 nodes, meaning that from 32 to 128 nodes the scaling is not linear anymore. This behaviour appears also for the \textit{Large} tests and it will be discussed further at the end of this Section.
\par For the \textit{Large} case (see Tab.~\ref{table:large_times}), the CPU tests are pure MPI, running on 32, 64, 128 nodes with 1024, 2048, 4096 MPI tasks, respectively (meaning 32 MPI tasks per node). Gridding time scales linearly when we double the computing resources, due to the fact that no communication overhead is included. For the GPU tests the gridding time does not scale increasing the number of accelerators from 128 to 512, remaining approximately constant. This trend is similar to what was found in the \textit{Intermediate} tests when more than 16 nodes were used. This behaviour is due to the overhead related to CUDA GPU memory management, and, more specifically, to repeated \textit{cudaMalloc} and \textit{cudaFree} calls that are implied by the iteration procedure through mesh sectors. At each iteration, the GPU memory storing visibilities is allocated and deallocated, causing the observed overhead. The number of CUDA calls increases with the number of sectors, which is equal to the number of MPI processes. With 2 GPUs these calls contribute to $\sim 20\%$ of the gridding time, while this number increases to $\sim 34\%$ in the 4 GPUs case. When large number of resources are used, it dominates the entire gridding time.
\begin{table*}[t]
\begin{center}
\centering \tabcolsep 1pt
\begin{tabular}{l|c|c|c|c|c|c|c|c}
\hline
 &  Nodes & MPI tasks (threads per task) & GPUs & Gridding (s) & Reduce (s) & FFT (s) & $w$-correction (s) & Total (s)\\ 
\hline
 & 32 & 1024 (1) & 0 & 4.5 & 9631.4 & 160.6 & 7.2 & 10246.0\\
CPU tests & 64 & 2048 (1) & 0 & 1.9 & 9598.2 & 107.1 & 3.5 & 10153.5\\
 & 128 & 4096 (1) & 0 & 1.1 & 9715.8 & 98.4 & 1.7 & 10266.5\\
\hline
 & 32 & 128 (8) & 128 & 2.6 & 54.8 & 4.2 & 0.3 & 67.4\\
GPU tests & 64 & 256 (8) & 256 & 2.4 & 59.4 & 2.8 & 0.2 & 69.4\\
 & 128 & 512 (8) & 512 & 2.7 & 72.6 & 2.7 & 0.1 & 83.4\\
\end{tabular}
\end{center}
\caption{Strong scaling results for each step of the code for \textit{Large} test, which has a mesh size of $65,536\times65,536\times32$ using LOFAR ILT dataset. Both full CPU and full GPU execution times in seconds are shown, with increasing computational resources for each. For these measurements we did not report the errors because multiple execution of this test would have consumed a large fraction of our available computing time.}
\label{table:large_times}
\end{table*}
\par \noindent{\bf FFT time.} While on the CPU the Fourier Transform is calculated using the FFTW, on the GPU the cuFFTMp library is adopted. In the \textit{Small} test (Fig.~\ref{fig:speedup_plots}, top right panel), the speed-up shows a peculiar behaviour. Within a single node, the performance worsens from 1 to 2 GPUs. This is because the FFT accounts for two distinct contributions: plan creation and actual computation. Due to the small mesh size, the plan creation contribution is comparable to computation causing the measured performance loss. From 2 to 4 GPUs the FFT scales linearly, exploiting the intra-node NVLink high-speed interconnect. On multiple nodes the FFT performance drops due to the progressively smaller problem size per GPU and the heavier communication overhead associated with the all-to-all required by the Fast Fourier Transform. Similarly, in the \textit{Intermediate} case (Fig.~\ref{fig:speedup_plots}, bottom right panel) the FFT scales efficiently up to 8 nodes, decreasing when a larger number of GPUs is increasingly used. In addition, the cuFFTMp has a startup time for initialising NVSHMEM and the 2D plan for the Fourier transform. This startup time ranges from 0.5~s to 1.5~s for the Intermediate test, increasing with the number of MPI tasks.
\par For the \textit{Small} tests on the CPU (Fig.~\ref{fig:speedup_plots}, top left panel), the FFT is computed through the hybrid FFTW MPI+OpenMP approach, keeping 8 OpenMP threads for each combination of MPI tasks. This explains the lack of scaling from 1 to 2 MPI tasks, where communication arising from calling multiple tasks determines an increase in the FFT time. The scaling is instead linear from 2 to 16 MPI tasks. In the CPU \textit{Intermediate} test (bottom left panel in Fig.~\ref{fig:speedup_plots}), the scaling is close to ideal up to 16 nodes. In Fig.~7 of Paper I, it turns out that in the pure MPI case the FFTW scaling reproduces the expected linear scaling up to $\sim 10^2$ MPI tasks. Here the hybrid MPI+OpenMP FFTW is used instead. This choice is due to the fact that we have less GPUs than CPU cores in each node, and to use efficiently the entire node OpenMP parallelization is unavoidable. Hybrid FFTW is faster than the MPI one when less cores than available are utilised, but at least in these tests that we have performed, the code scalability gets worse with 128 MPI tasks distributed among 32 nodes. Understanding this scaling of the FFTW with OpenMP requires a deep code's profiling and it is beyond the scope of this paper.
\par For the \textit{Large} case (Tab.~\ref{table:large_times}), in pure CPU tests the FFT's performance improves by $\sim 60\%$ when passing from 32 to 64 nodes, while the performance gain is poor when passing from 64 to 128 nodes. With such a large number of MPI tasks, all to all communication dominates the runtime and the performance gain in the computation is almost completely suppressed. In the GPU tests, the FFT's performance improves by $\sim 50\%$ when passing from 32 to 64 nodes, while the performance gain drops when using 128 nodes. Again, this is due to the all-to-all communications and, at the same time, the progressively lower computation related to an increasingly larger amount of MPI processes.
\par \noindent{\bf $w$-correction time} Strong scaling tests of the $w$-correction reveal that the code scales almost linearly both for the \textit{Small}, hybrid MPI+OpenMP, CPU tests and for the \textit{Intermediate}, pure MPI, CPU tests. The same holds for GPUs, for which this step is implemented within a CUDA kernel. For this algorithm GPUs are more than order of magnitude faster than CPUs. In the \textit{Large} tests, $w$-correction time keeps scaling linearly in both CPU and GPU tests. In particular, this kernel turns out to run efficiently on GPUs, there existing a factor of $\sim 20$ on average in runtime between CPU and GPU tests. $w$-correction is performed directly on the grid which has already been mapped on GPUs, such that no unnecessary memory management is needed.
\par \noindent{\bf Full code}
Overall, RICK shows a sub-linear scalability in both the \textit{Small} and \textit{Intermediate} configurations, both for the CPU and the GPU tests. As already pointed out above, such behaviour can be mainly ascribed to the impact on the code of the reduce related overhead and is also evident from the figures presented in Tab.~\ref{table:large_times}. Additional factors, like the progressively smaller amount of data to process as the number of computing units grows, affect in particular the GPU runs.

\subsection{Weak scaling tests}
\begin{figure*}[ht!]
\centering
\includegraphics[width=\textwidth]{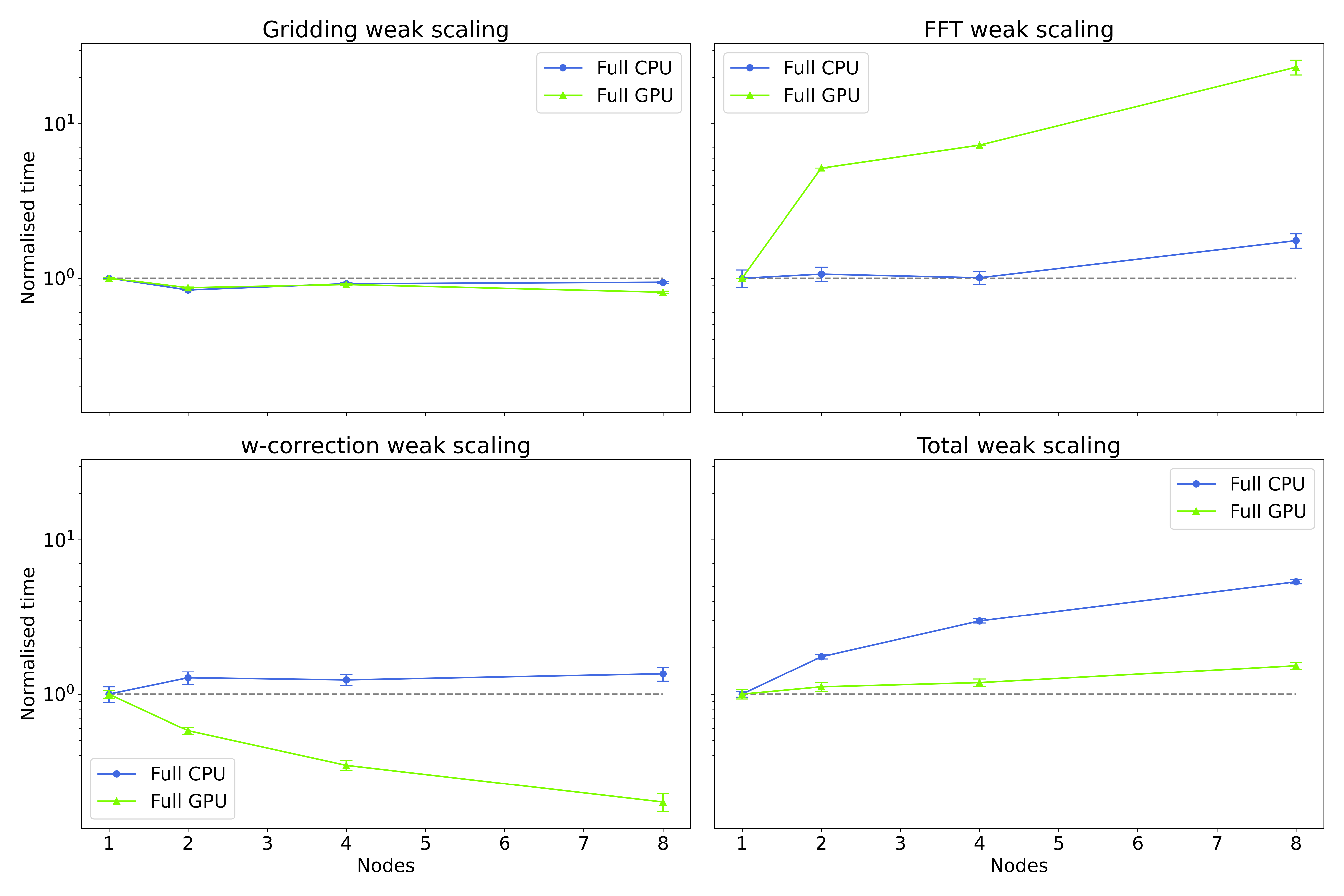}
\caption{Results for the weak scaling tests for each step of the code. The \enquote{normalised time} on the $y$ axis is defined as the ratio between the wall clock time measured using $N_{p}$ computing nodes for a $N_{1} \times N_{p}$ size problem and the wall clock time measured using a single computing node for a $N_{1}$ size problem (which is our baseline configuration). Values below 1 indicate an enhancement in performance, whereas values above 1 represent a reduction in scalability. The horizontal dashed grey line represents the ideal weak scaling of the code. From the top left corner, clockwise, results are shown for the gridding, FFT, $w$-correction, and total. The full CPU execution is shown in blue, while in green the full GPU enabling of the code. For each run 8 OpenMP threads were spawned.}
\label{fig:weak_scaling_tests}
\end{figure*}
Weak scaling tests have been performed using the LOFAR Dutch dataset over $1$, $2$, $4$ and $8$ computing nodes, instancing 4 MPI tasks per node and doubling at each step both the input size data and the number of $w$ planes. For each configuration, 8 OpenMP threads per MPI task have been spawned. In Fig.~\ref{fig:weak_scaling_tests}, we present the runtime normalised to that of a single node (adopted as reference) of the different parts of the code and of the full code. Ideal weak scaling would result in a value of 1 regardless of the number of nodes.
\par Gridding, top-left panel, shows an almost ideal weak scalability for both the CPU and the GPU tests, with performance slightly improving increasing the number of nodes from 1 to 2 and showing small fluctuations for 4 and 8 nodes. 
The FFT weak scaling is presented in the top right panel. The hybrid MPI+OpenMP FFTW3 scales almost ideally from 1 to 4 nodes. On 8 nodes, the performance loss can be attributed to the impact of the all-to-all communication. For the GPU, cuFFTMp based tests, a significant performance drop is observed when the number of nodes increases from 1 to 2. This is mainly due to the switch from the intra-node high-speed NVlink interconnect to the lower bandwidth Infiniband network connecting different nodes. From 2 to 8 nodes, the performance is also worsening. This can be explained as a result of the way the weak scaling test is designed, maintaining constant the grid dimension and doubling at each step the number of $w$-planes. This increases the number of FFTs performed by the algorithm. However, the amount of computation per FFT that is performed by each GPU decreases with increasing the number of GPUs, leading to a computing efficiency loss, reflected by the weak scaling curve.
\par In the bottom-left panel, the weak scaling of the $w$-correction term is presented. The weak scaling for the CPU is nearly optimal, whereas in the case of the GPU, performance tends to increase significantly as the number of nodes grows. This behaviour can be interpreted as a consequence of the decreasing volume of data transferred from the GPU to the CPU as the number of GPUs increases. Specifically, when the number of GPUs is doubled, the portion of the image that needs to be copied back to the host halves. This has a positive impact on the time needed to transfer data between the device and the host. 
\par When considering the overall scalability of the code, GPUs exhibit just a small reduction in efficiency as the number of nodes increases. In the case of the CPU instead, the deviation from the optimal scaling is evident. Once more, the efficiency loss is mainly due to the effect of the reduce, as we will discuss in the next section.

\section{Discussion}
\label{sec:reduce}

The primary goal of the RICK code is to efficiently process huge datasets and  generate large images in a reasonable time scale, of the order of seconds or minutes. In the tests provided, we utilised datasets with visibilities of up to 533 GB in size. Nevertheless, datasets of any size can be easily managed by splitting them into frequencies or time chunks. Chunks can be loaded sequentially, as described in~\citet{Gheller2024HPC}. This also enables to reserve an appropriate fraction of memory for the computational mesh.
\par By employing an appropriate number of processing units, we have successfully generated images with a resolution of $65,536^2$ pixels. Utilising GPUs significantly enhances the performance of all code components, resulting in a considerable decrease in the time to solution. In particular, the exploitation of the cuFFTMp library reduces the impact of the Fourier Transform step. The corresponding computing time gain can range from one to two orders of magnitude compared to the FFTW CPU-only approach, depending on the scale of the problem. The current implementation of the library is highly efficient within a single node, taking advantage of the NVlink interconnect. However, its scalability is limited when multiple nodes are used due to the slower network and the use of the NVSHMEM protocol. In the case of the FFT, this protocol does not seem to offer optimal scaling on a large number of GPUs, unlike other solutions such as NCCL, which is used for the gridding part.
In all the CPU tests, the hybrid MPI+OpenMP version of the FFTW ensures good scalability. However, a slight performance loss is found in a few configurations. This cannot be easily explained, but it is interesting noting that this issue has not been observed using the MPI-only version of the library.
\par Our tests have demonstrated that, when the problem size and computational setup are increased, the code's performance and scalability are significantly impacted by MPI communication. In addition to the parallel FFT, communication is performed by the reduce operations required to collect and add up the mesh data from all computing units during the gridding step. This last contribution tends to to dominate increasing the number of computing units. The remainder of this Section discusses the details of this aspect.

\begin{figure}[t]
\subfloat{\includegraphics[width=\columnwidth]{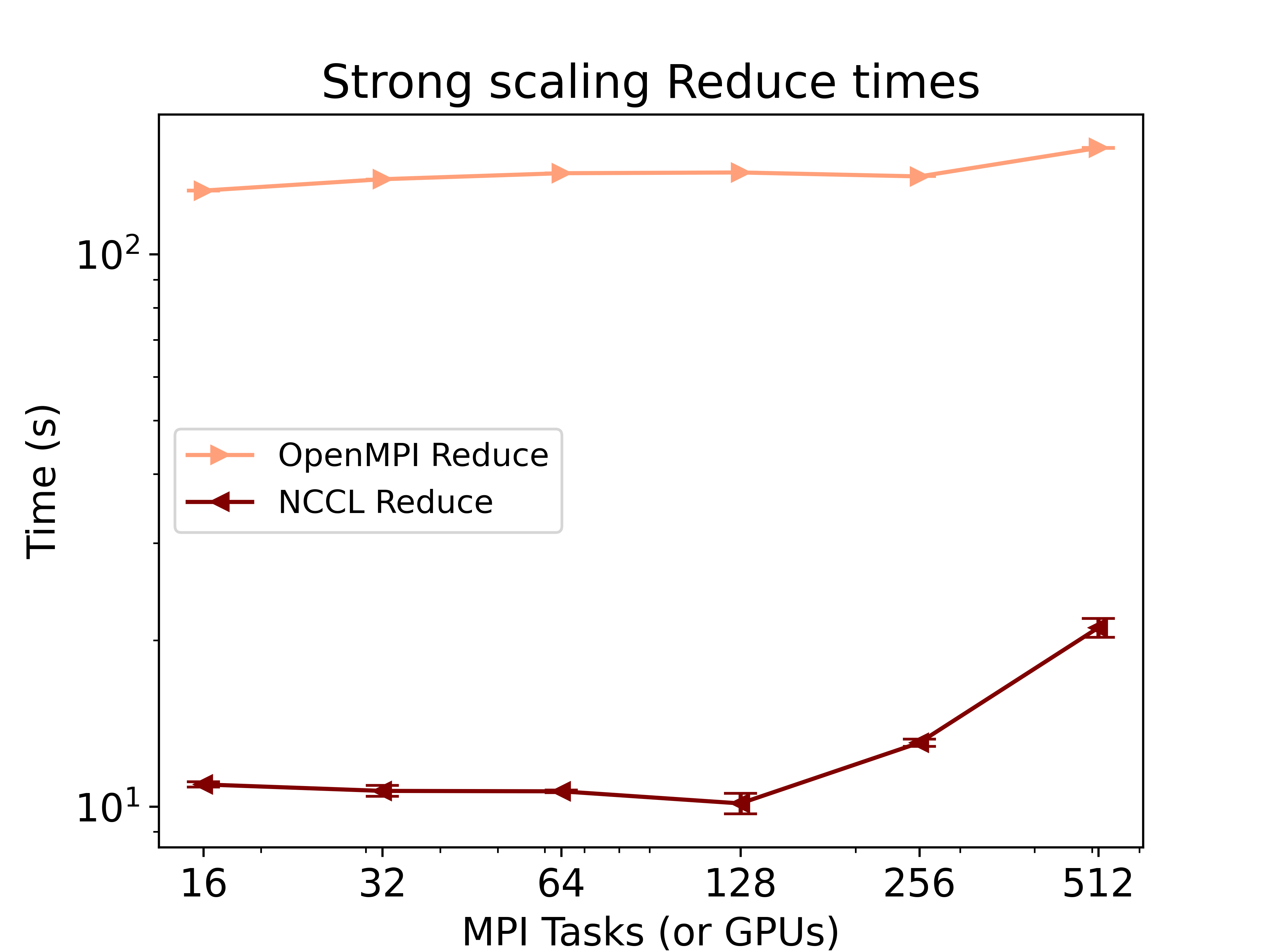}}\\
\subfloat{\includegraphics[width=\columnwidth]{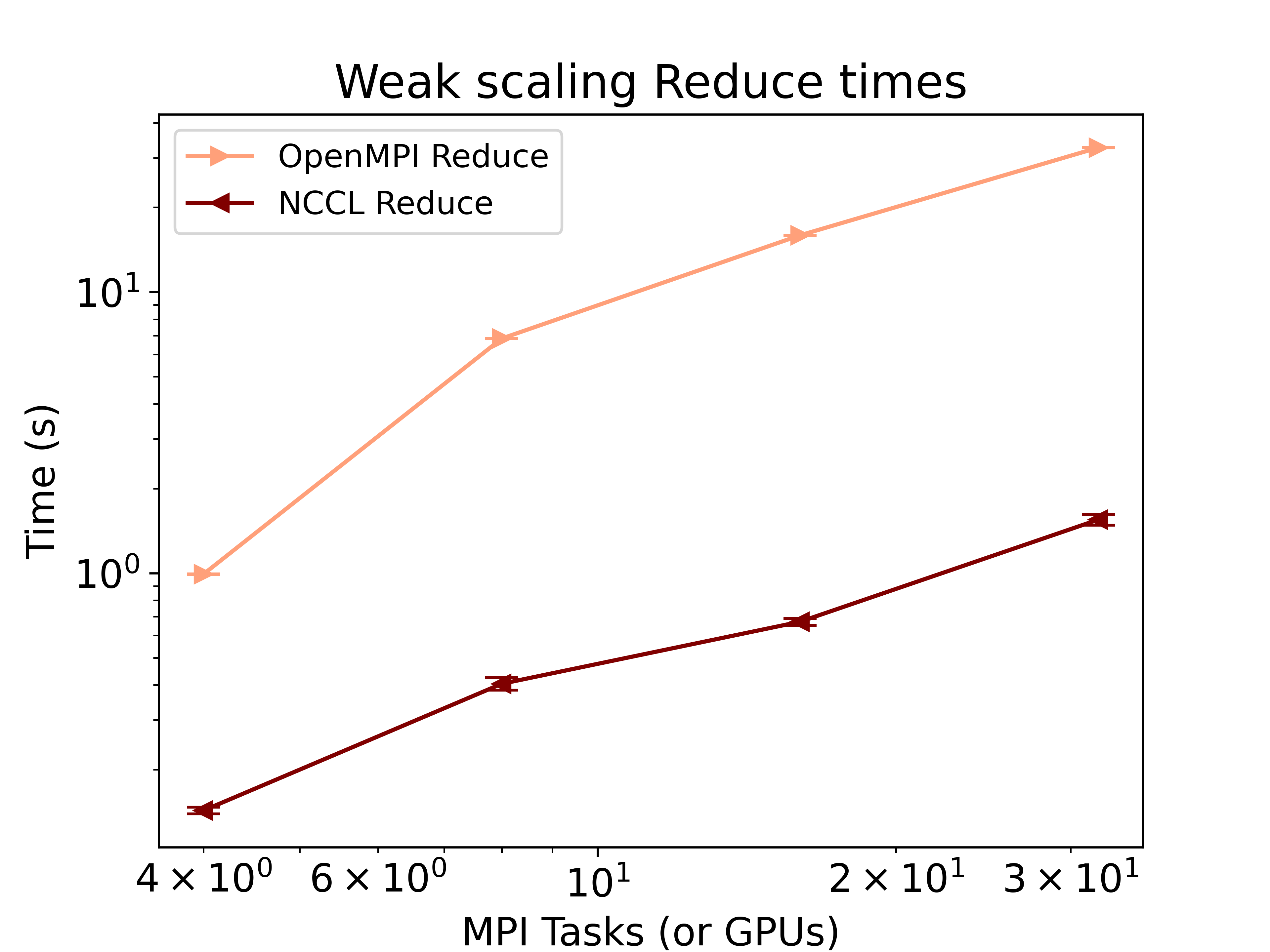}}
\caption{Time results for the reduce operation versus the number of MPI tasks or GPUs for the \textit{Intermediate} strong scaling test (top panel), and the weak scaling test (bottom panel). Comparison between the OpenMPI Ireduce (in pink) and NCCL GPU reduce (in brown).}
\label{fig:reduce_time}
\end{figure}

\par Visibilities are read evenly in parallel from an input data file in a time-log order, i.e. if we are dealing with an $8$ hours observation and $N_{\rm{p}} = 8$, each MPI task reads data for one hour of observation. Conversely, the Cartesian computational mesh, where visibilities are gridded, is divided into rectangular slabs in the $u$-$v$ plane. Each slab is allocated to a distinct MPI task. Therefore, it is necessary to convert time-ordered into space-ordered data. To achieve this, the code iterates through the slabs. During each iteration $i$, each MPI task simultaneously computes its contribution to the $i$-th slab. Then, each MPI task stores its local contribution in an auxiliary buffer. Once local contributions are computed, they are summed together on a {\it target} buffer located on the $i$-th MPI task. Data collection and aggregation are achieved using an MPI reduction operation, which combines both communication (collecting data from several sources) and computation (calculating the sum of all contributions). As discussed in Paper I, for large images the code spends $80-90\%$ of the total runtime in the reduce operation, when the reduce operation was implemented with standard Ireduce from OpenMPI library.
\par The overhead introduced in the code by the reduction operation depends on several factors, namely: {\it i)} the amount of data communicated among the MPI tasks; {\it ii)} the number of MPI reduce calls {\it iii)} the amount of computation required to perform the sum; \textit{iv)} the network topology. These factors are interconnected. During our iterative procedure, the number of iterations over slabs is equal to $N_{\rm{p}}$. Each MPI task performs a reduce at every iteration, hence 
the theoretical reduce time $T_{\rm{R}}$ can be estimated as the sum of the communication ($T_{\rm{co}}$) and the calculation ($T_{\rm{ca}}$) times. For the CPU implementation, using the OpenMPI library, $T_{\rm{R}}$ can be calculated as: 
\begin{equation}
\begin{split}
    T_{\rm{R}} & = (T_{\rm{co}} + T_{\rm{ca}}) \times N_{\rm{p}} \\
    & = \Bigl(\beta \frac{D}{N_{\rm{p}}} + \lambda \log{N_{\rm{p}}}\Bigr)\times N_{\rm{p}} + \Bigl(\frac{D}{N_{\rm{p}}} t_{\rm{sum}}\Bigr) \times N_{\rm{p}}  \\
    & = (\beta + t_{\rm{sum}})D + \lambda N_{\rm{p}} \log{N_{\rm{p}}}~,
\end{split}
\end{equation}
where $D$ is the total size of the data, $\beta$ quantifies the bandwidth and the network topology, $\lambda$ is the network latency, and $t_{\rm{sum}}$ is the time required for a single summation operation. The logarithmic term is due to the logarithmic tree algorithm used by the OpenMPI reduce.
\par The reduce time is then composed by two terms: the first depends on the data size and on $\beta$, while the second on the number of MPI processes. The latency term $\lambda$ is generally very low ($\leq 0.6~\rm{\mu s}$). In the strong scaling case, in which $D$ is maintained constant increasing the number of MPI tasks, the first term is dominant on the second because of the low latency: for this reason, we observe a constant reduce time (Fig.~\ref{fig:reduce_time}, top panel). Only once we use a considerably large $N_{\rm{p}}$, e.g. over 128 MPI tasks (as reported in the top panel of Fig.~\ref{fig:reduce_time}), we start to observe an increase in the reduce time: however, such increase is not due to the processes themselves, but rather to the network topology, because of the non-ideal interconnection between the nodes and to the impact of the latency term. For the weak scaling, instead, we have $D$ that increases as well as $N_{\rm{p}}$. The first term is still dominating, and for this reason we observe a reduce time which increases linearly with the data size (Fig.~\ref{fig:reduce_time}, bottom panel).
\par The GPU implementation exploits the NCCL library, that implements a ring algorithm for the reduce: this means that $T_{\rm{R}}$ can be written as
\begin{equation}
T_{R} = (\beta + t_{\rm{sum}})D + \lambda N_{\rm{p}}^2~.
\end{equation}
Due to the quadratic dependency from $N_{\rm{p}}$, latency becomes non-negligible using a smaller amount of processes. For example, OpenMPI would necessitate 100,000 MPI tasks to produce a latency of $1$ second, whereas the NCCL would require only 1,000 MPI tasks. On the other hand, GPUs are capable of completing the sum operation much faster than the CPUs, because of their large computational throughput, about two order of magnitude larger than that of the CPU. This results in a considerably smaller $t_{\rm{sum}}$, that contributes in reducing the NCCL reduce time. In addition the NCCL reduce performs better than the MPI one due to \textit{i)} the high-speed NVlink interconnection for intra-node GPU-GPU communication, \textit{ii)} the number of network interface cards (NIC) that equals the number of accelerators per node, leading to a bandwidth 4 times bigger than pure CPU cases. Overall we see that the the reduce operation is about $20$ times faster on the GPU for the \textit{Small} test (see Tab.~\ref{table:best_times}) and up to $\times 175$ faster for the \textit{Large} tests (see Tab.~\ref{table:large_times}).
\par Tab.~\ref{table:large_times} also shows that the MPI Ireduce time remains constant scaling from 32 to 64 to 128 computing nodes. This has a progressively greater impact on the code compared to the more computationally intensive components of the algorithm. The GPU reduce time tends even to increase with the number of nodes. However, it is two orders of magnitude lower than that of the CPU. In both scenarios, increasing the number of nodes does not result in any improvement in speed, and in the case of the GPU, it actually leads to a decrease in performance. 
\par In general, in all test regimes the code runtime is strongly impacted by the reduce operation.  In order to exploit with the maximum efficiency the available computing systems, it is crucial to select an hardware configuration that has the fewest computing units necessary to meet the memory needs. This also results in saving energy, since the problem is solved in approximately the same amount of time by while using less computational resources.

\section{Conclusions}
\label{sec:conclusions}
In this paper we presented RICK, a publicly available code which aims to address the imaging problem in radio astronomy using a HPC-based approach throughout parallelism and acceleration through GPUs. 
\par Our main achievements can be summarised as follows.
\begin{itemize}
    \item Starting from the code presented in~\citet{gheller2023}, we managed to complete the full GPU porting of the code, including the FFT operation and the communication. Currently the code is capable of fully running on NVIDIA GPUs: this minimises data movement between CPU and GPU and avoids overheads related to the CPU communication, exploiting the full potential of the GPU enabling.
    \item For the reduce operation we present an implementation for GPUs based on the NVIDIA NCCL library, which uses, where possible, high-speed interconnectivity. Moreover, we introduced an hybrid approach MPI+OpenMP to the reduce problem which has been used here for testing but will be detailed in a future paper.
    \item The FFT is instead accelerated using the cuFFTMp library of the NVIDIA HPC-SDK toolkit: with this the FFT problem can be distributed among multiple GPUs, which is critical considering the large volume of data processed compared to the memory of the devices. To exploit the full CPU computational capacity of a node we also tested a hybrid FFTW, which combines MPI tasks with OpenMP threads, which has a better scalability compared to the sole MPI FFTW thanks to a reduced communication overload.
    \item We tested performance and scalability of the code on Leonardo HPC cluster at CINECA (Italy), with tests on strong and weak scaling of every step that constitute our code using real LOFAR data (also comprising International stations). Comparing our full-GPU code with the full-CPU, we observe a gain in total runtime of a factor $\times9$ for the \textit{Small} and \textit{Intermediate} dataset configuration within a single node, and up to $\sim \times 130$ for the \textit{Large} configuration over a large number of nodes.
    \item Thanks to GPU offloading, our code's computational costs have been greatly reduced. The runtime is now mainly affected by communication. To fully leverage the improved capabilities of this implementation, it is essential to select a computational setup that match the problem size and and minimises communication as much as possible.
\end{itemize}

Overall, RICK stands out as an innovative radio imaging software that fully utilises GPUs, making it a promising solution for the future software suites for processing big radio astronomical data, such as those expected for the SKA. Future advancements will focus on optimising communication and providing full support for parallel I/O in radio astronomy measurement  sets.

\section*{Acknowledgements}
This paper is supported by the Fondazione ICSC, Spoke 3 Astrophysics and Cosmos Observations. National Recovery and Resilience Plan (Piano Nazionale di Ripresa e Resilienza, PNRR) Project ID CN\_00000013 "Italian Research Center for High-Performance Computing, Big Data and Quantum Computing" funded by MUR Missione 4 Componente 2 Investimento 1.4: Potenziamento strutture di ricerca e creazione di "campioni nazionali di R\&S (M4C2-19)" - Next Generation EU (NGEU), and it's also supported by (Programma Operativo Nazionale, PON), "Tematiche di Ricerca Green e dell'Innovazione". We acknowledge the CINECA award under the ISCRA initiative, for the availability of high performance computing resources and support. The HPC tests and benchmarks this work is based on, have been produced on the Leonardo Supercomputer at CINECA (Bologna, Italy) in the framework of the ISCRA programme IscrC\_RICK (project: HP10CDUNG6). The data to perform all the tests have been kindly provided by the LOFAR project LC14\_018 , PI F. Vazza, and project LT16\_005, PI A. Botteon.

\bibliographystyle{elsarticle-harv}
\bibliography{franco,franco2,bib_add}

\end{document}